# MATHEMATICAL EQUIVALENCY OF THE ETHER BASED GAVITATION THEORY OF JANOSSY AND GENERAL RELATIVITY

György SZONDY
E-mail: gyorgy_szondy@hotmail.com
V1.9 – 2003. October 23.

**Abstract**

*„There are several interpretations and approaches to relativity. All of them are characterized by the fact that none of them is accepted by physicists without doubts, even the Einsteinian General Relativity!*
*Only those theories can get into the spotlight that predicts something that is different from predictions of other concurrent theories.*
*From this point of view the theory of Janossy [1] is not an excellent idea, as he tried to show that his materialistic approach also corresponds to the general principles and equations of relativity.*
*As his program was basically successful there was not any additional result, except the philosophical part.."*
Since the death of Janossy, his work has almost been forgotten. Both what he achieved and what he was not succeeded in. He was one of the founders of KFKI (Central Physics Research Institute of the Hungarian Scientific Academy) but his effort has not been carried over even there, however his collegues are still remember his name and his work. Luckily his books are still available in the Hungarian libraries. Additionally the most informed etherists in the world who are lucky enough to know his work are consider his work as a No1 reference.
This paper is designated to refresh the idea of the ether based gravitation theory of Janossy and introduce a well founded way to adjust it to be equivalent with the experiences and General Relativity.

## 1 Introduction

In his book „Theory of Relativity according to the physical reality" that was published in 1973 Janossy showed that the relativistic effects described in General Relativity can also be described on euclidean space by Ether theory. He described the gravity by the optical property of ether, more precisely the dependency of the speed of light on the gravitation potential.
His results were qualitatively good, but in case of gravitational light deflection and the relativistic part of the Mercury perihelion advance they were half of the expected value.

We found that by reviewing his assumptions and correcting the definitions we can also correct these errors and create a description that is mathematically equivalent with the Einsteinian one.

## 1. Result and errors in the description of Janossy

Janossy described experiences beyond the Newtonian gravity differently from Einstein. He created a model on Euclidean space and applied optic. He explained the gravitation light deflection and the relativistic perihelion advance of Mercury based on the potential dependency of the speed of light.
In the introduction of his theory Janossy considers three relativistic effects; gravitation light deflection and the relativistic perihelion advance of Mercury and the gravitational red-shift.
Calculations are based on the following metric:

$$g(x) = \begin{bmatrix} 1 & 0 & 0 & 0 \\ 0 & 1 & 0 & 0 \\ 0 & 0 & 1 & 0 \\ 0 & 0 & 0 & -c^2(r) \end{bmatrix} \quad (1)$$

, where c(r) refers to the position dependency of the speed of light.
The metric above defines the connection between the gravitational potential and the speed of light. From this position dependency a value for light deflection and also for the relativistic part of the Mercury perihelion shift can be calculated. The order of the result is correct in both case.
Upon these results Janossy concludes that Ether based description of the gravitation is qualitatively good.

Unfortunately all the results – except the gravitational red-shift that was defined by $g_{44}(r)$ – are not accurate.
Assuming that the form of the metric is (1) [1-p367.18], where

$$c(r)^2 = \text{constant} + 2\theta(r) \qquad (2)$$

[1-p367.21] and $\theta(r)$ is the gravitational potential, value calculated for the light deflection [1-p372.37b] is exactly the half of the measured and the Einsteinian correction. The value calculated by Janossy for the relativistic perihelion advance was

$$d\varphi = 4\pi \frac{v^2}{c^2} \qquad (3)$$

The Einsteinian value that matches the experiments is

$$d\varphi = 6\pi \frac{v^2}{c^2} \qquad (4)$$

while Newtonian gravity with velocity dependent mass gives

$$d\varphi = 2\pi \frac{v^2}{c^2} \qquad (5)$$

It seems that (3) that contains coordinate deflection part is right between (4) and (5), therefore we can state that the coordinate deflection part in the perihelion advance is also the half of the right value.
Janossy also mention that the proper value can be obtained from the theory by the sufficient change in the definition regarding to `c(r)`$^2$, but he considered this modification would not be well founded in the current state [1-p371]. He stated that any additional modification should be based on a well founded change in the correlation of the metric and the gravitational potential.

Let us notice that in case of the metric above the gravitational redshift is

$$\nu_1 / \nu_2 = c(r_1) / c(r_2) \qquad (6)$$

[1-p372.39a]. It also defines the size of the particles being constant: independent of the gravitational potential.

## 2. Proposed corrections in the theory

The assumption that the size of the particle is independent from the gravitational potential is really useful from the perspective of the mathematical complexity but is not well founded restriction.
However these kind of restrictions are more or less usual. For example in 1961 Brans and Dicke [3] dealt with the problem why we consider the rest mass of the particle being independent of the position in General Relativity.
In case of the theory of Janossy it would be better to define the gravitational red-shift with the

$$\nu_1 d_1 / \nu_2 d_2 = c(r_1) / c(r_2) \qquad (7)$$

equation, where besides the atomic frequency the atomic size (`d`) may also change.

The proper link between the change of frequency, atomic size and the speed of light can be determined in at least two ways:
The easiest one if we consider the optical behavior of the light and conclude that in case of weak fields the change in the speed of light is twice as much as it was assumed by Janossy. Unfortunately it does not define the metric properly, additionally it does not explain the change at all.

Second approach – described below – is based on the assumption that physical laws are invariant:

Let us consider an object having spherically symmetric gravitation field and mass **M**. We watch this object from two points defined by `r₁` and `r₂`. Physical laws assumed being invariant at each points, therefore we are allowed to determine the transformation rule between the two points using the Einsteinian General Relativity.

Speed of light at **r₁** is c(**r₁**), the mass of the object is M₁ the local atomic frequency is $v_1$ and the local size of the atom is d₁. These values at **r₂** are measured to be c(**r₂**), M₂, $v_2$ and d₂.

Let us assume that the gravitational red-shift between the two points is

$$v_1 / v_2 = \alpha \tag{8}$$

Then the observed mass of the body at r₂ is

$$M_2 = M_1 / \alpha \tag{9}$$

, because the mass scale changes with the atomic frequency: we experience inverse of the mass defect.

The Schwarzschild radius of the body at r₁ using d₁ as a unit is

$$r_{g1} = 2M_1 G / c_0^2 = N d_1 \tag{10}$$

Where c₀ is the invariant speed of light. The Schwarzschild radius at r₂ substituting d₂ and M₂ from equation (9) and (10)

$$r_{g2} = 2M_1 G / \alpha c_0^2 = N d_1 / \alpha \tag{11}$$

The two values are certainly the same, therefore we get

$$d_2 / d_1 = \alpha \tag{12}$$

Multiplying the reciprocal value of (8) and the reciprocal value of (12) we get the left side of the equation (7)

$$v_1 d_1 / v_2 d_2 = 1 / \alpha^2 \tag{13}$$

We got a new equation for the correlation of the gravitational red-shift and the speed of light that replaces equation (6)

$$v_1 / v_2 = \sqrt{c(r_1) / c(r_2)} \tag{14}$$

Considering this new result we get a different metric instead of (1)

$$g(x) = \begin{bmatrix} \frac{c_0}{c(r)} & 0 & 0 & 0 \\ 0 & \frac{c_0}{c(r)} & 0 & 0 \\ 0 & 0 & \frac{c_0}{c(r)} & 0 \\ 0 & 0 & 0 & -c_0 c(r) \end{bmatrix} \tag{15}$$

It means that particles in regions with lower speed of light are receive equivalent change in the atomic frequency and the atomic size. Let us remark that the form of the new metric is the same as the conform-euclidean approximation of the Schwarzschild metric.

3. **Results of the changes**

It is obvious that after our change we received double change in the speed of light for the same gravitational red-shift than Janossy. Therefore light deflection will be the proper value – twice as much as calculated originally by him.

Calculations regarding to the perihelion advance is more complex, therefore we have not applied it here. However we assume that the part that is caused by the relativistic change of the mass will be the same and the part caused by the deformation of the light-based coordinate system will be doubled. With this we assumed that this value will also be the same as the Einsteinian.

One of the achievements of the change that in this ether based description the relativistic corrections are the same as calculated using the Einsteinian approach. Besides – as $g_{kk}$ (k = 1,2,3) are also changing – the size of particle is also changing that causes not only additional mathematical complexity, but predicts a possible additional tidal effect during the free fall.

### 3.1. Cosmological results

Just like in Linear Brans-Dicke gravity [4] during the collapse of a spherical object there is a critical size near the Scwarzschild radius, where the inner region apparently inflates This inflation is caused by the reduction of light-speed. This inflation means that approaching to the center of the object we measure bigger radius; in this region the meaning of inside and outside is inverted.
The inside region can be interpreted as a new "inner" Universe. For this reason reaching the critical size for a collapsing start means a Big Bang for the inner Universe.

**Note**: I found recently that this idea has already been introduced in the 90s by Lee Smolin. It has been mentioned for example in a lecture of Woehler. [6]

## 4. Summary

We concluded that the ether based modified Janossy description of gravitation is also usable for calculating relativistic effects of the gravitation. There might be new predictions to validate, additionally as the new description is based on Euclidean geometry the mathematical complexity is reduced. The description also answers certain cosmological questions.

In this article we defined the way of correction to be applied for the description of Janossy to receive the proper value for the gravitational light deflection. Also the change in results regarding to the perihelion advance is as expected both in tendency and estimated value. Therefore we consider this approach being promising.
The modified ether-based gravitation theory of Janossy as an approach equivalent to the Einsteinian is remarkable.

## 5. References

[1] Lajos Jánossy, *Theory of Relativity according to the physical reality* (*Relativitás Elmélet a fizikai valóság alapján)*, 1973 Akadémia Kiadó, Budapest
[2] Péter Hraskó, *Theory of Relativity (Relativitás Elmélet)*, 2002 Typotex Kiadó
[3] C. Brans and R. H. Dicke, *Mach's Principle and a Relativistiv Theory of Gravitation*, Phys. Rev. D 124-925
[4] György Szondy, *Linear Relativity as the Result of Unit Transformation*, physics/0109038
[5] Landau -Lifsiz, *Theoretical Physics II – Classical Fields*, 1973
[6] Dr. Kai Woehler, *THE MULTIVERSE*, http://www.mira.org/bonestell/kai/multi.htm